\documentclass[journal,article,submit,moreauthors]{mdpi} 
\UseRawInputEncoding
\firstpage{1} 
\hreflink{https://doi.org/} % If needed use \linebreak

\usepackage{wrapfig}
\usepackage[pscoord]{eso-pic}
\usepackage[fulladjust]{marginnote}
\usepackage{tikz}
\usepackage[many]{tcolorbox}
\tcbuselibrary{skins, breakable, theorems}

\usepackage{varwidth}
\usepackage{extarrows}
\usepackage{tikz}
\usepackage{tabu}
\usetikzlibrary{arrows}
\usepackage{tcolorbox}
\tcbuselibrary{skins, breakable, theorems}
\usepackage{fancybox}
\usepackage{amsthm}

\reversemarginpar

\definecolor{Gray}{gray}{.25}
\definecolor{lightgray1}{gray}{0.95}
\definecolor{lightgray2}{gray}{0.6}
\definecolor{lightgray3}{gray}{0.8}
\Title{Energy spectrum of interacting gas: cluster expansion method}

\Author{Hao-Dan Li $^{1}$, Shi-Lin Li $^{1}$ , Yu-Jie Chen $^{1}$ , Wen-Du Li $^{2,\ddagger}$and Wu-Sheng Dai $^{1,\ddagger}$}

\address{%
$^{1}$ \quad Department of Physics, Tianjin University, Tianjin 300350, P.R. China\\
$^{2}$ \quad College of Physics and Materials Science, Tianjin Normal University, Tianjin 300387, PR China
}

\firstnote{liwendu@tjnu.edu.cn.} 
\secondnote{daiwusheng@tju.edu.cn.}
\abstract{In this paper, we calculate the energy spectrum of interacting gases by
converting the cluster expansion method in statistical mechanics into a method
of solving energy eigenvalues. We obtain an explicit expression of the energy
eigenvalue, by which we can calculate the eigenvalue of an interacting gas
from the interparticle potential directly. As an example, we calculate the
energy spectrum for an interacting gas with soft-sphere potentials.}


\usepackage{titletoc}   
\titlecontents{section}[0.6em]{\vspace{.25\baselineskip}}
             {\thecontentslabel\quad}{}
             {\hspace{.5em}\titlerule*[10pt]{$\cdot$}\contentspage}
\titlecontents{subsection}[2.1em]{\vspace{.25\baselineskip}}
             {\thecontentslabel\quad}{}
             {\hspace{.5em}\titlerule*[10pt]{$\cdot$}\contentspage}
\normalfont \ttfamily
\numberwithin{equation}{section}
\begin{document}
%%%%%%%%%%%%%%%%%%%%%%%%%%%%%%%%%%%%%%%%%%
\nolinenumbers

% now start line numbers
%\linenumbers

% the * after section prevents numbering
%\section*{Introduction}

\section{Introduction}

The energy spectrum of an interacting many-body system, in principle, can be
obtained by solving the eigenequation of the Hamiltonian $H$:%
\begin{equation}
H\psi=E\psi, \label{eigeneq}%
\end{equation}
where $E$ is the energy eigenvalue and $\psi$ is the eigenfunction. However,
it is very difficult to solve the eigenequation\ (\ref{eigeneq}) for
interacting many-body systems. In this paper, instead of solving
eigenequations, we solve the energy spectrum based on the cluster expansion in
statistical mechanics. This approach comes from an observation of the
partition function. The partition function is defined as%
\begin{equation}
Z\left(  \beta\right)  =\sum_{s}e^{-\beta E_{s}}, \label{Zbeta}%
\end{equation}
where $\beta=\frac{1}{kT}$ and the sum runs over all states. If the energy
eigenvalue $E_{s}$ is known, the partition function $Z\left(  \beta\right)  $
can be obtained by performing the sum in Eq. (\ref{Zbeta}). However, if what
we know is the partition function rather than the energy eigenvalue, then Eq.
(\ref{Zbeta}) is a sum (or, equivalently, an integral) equation of
eigenvalues. This inspires us that when the partition function is known, we
can seek the energy eigenvalue by solving the sum (an integral) equation
(\ref{Zbeta}). In statistical mechanics, many methods have been developed for
calculating partition functions, which do not need to know the energy
eigenvalues in advance. Therefore, we can first work out the partition
function in statistical mechanics, and then solve the energy eigenvalue by
solving the sum (integral) equation (\ref{Zbeta}).

In statistical mechanics, mechanical quantities are replaced by their average
values. Taking average, of course, loses information. Let us see what
information is lost and what information is retained after averaging. The
definition of the partition function (\ref{Zbeta}) contains only eigenvalues,
but no eigenfunctions. That is, after the statistical average, the
eigenfunction information is lost, but the eigenvalue information is retained.
It is worth noting that all the eigenvalue information is retained in the
partition function. This is because although the partition function is an
average value, it is the average value of all temperatures, i.e., the average
value is a function of temperature. In other words, the partition function is
not a single average value, but an infinite number of average values, as
informative as all the eigenvalues. All thermodynamic quantities can be
calculated from the partition function, so the thermodynamic quantity also
contains eigenvalue information. In this paper, we take the partition function
as an example to illustrate the method. The method is also applicable to
various thermodynamic quantities.

The method directly solves the eigenvalue from the partition function, without
solving the eigenfunction at the same time. Seeking a method for solving
eigenvalues without solving the eigenfunctions is important in quantum
mechanics. Recall that in quantum mechanics, in most cases, even if we do not
need the eigenfunction, we still usually have to solve the eigenfunction at
the same time when solving the eigenvalue. For example, in perturbation
theory, in order to solve the $n$-th order perturbation of eigenvalues, we
must first solve the $\left(  n-1\right)  $-th order perturbation of
eigenfunctions. The partition function contains the information of
eigenvalues, but the information of the eigenfunction has been averaged out.
Therefore, only the eigenvalue can be extracted from the partition function.
In fact, this is just the advantage of this method: solving eigenvalues
without solving eigenfunctions at the same time. In statistical mechanics,
many methods for calculating partition functions have been developed, which
calculate the partition function without knowing the eigenvalues in advance,
such as the cluster expansion method. Our method converts a method for
calculating partition functions and thermodynamic quantities in statistical
mechanics into a method of calculating the energy eigenvalue of interacting
gas systems.

In this paper, using the method, we find an explicit expression for the energy
eigenvalue of interacting gases. By this result, we can calculate the energy
eigenvalue from the interparticle potential directly.

The method is based on the canonical partition function. As long as the
canonical partition function is known, the energy eigenvalue spectrum can be
calculated using the method. Many methods for calculating the canonical
partition function are developed
\cite{chaturvedi1996canonical,deldar2016partition,hsieh2016efficient,do2011rapid,lee2012exact,sisman2021fractional,zhao2020bose}%
, such as the cluster expansion
\cite{pulvirenti2012cluster,wang2016derivation,bastianello2016cluster,fronczak2013cluster,drautz2006obtaining}%
. Canonical partition functions are also calculated through the virial
expansion \cite{bannur2015virial,ushcats2013adequacy,siudem2013partition}. The
eigenvalue spectrum of interacting many-body systems is an important problem
and many methods are developed
\cite{settino2020emergence,pachos2018quantifying,faldella2014complete,christandl2014eigenvalue}%
, such as classical limit of the quantum-mechanical canonical partition
function \cite{seglar2013classical}, the effect of particle-wall interactions
on gases \cite{firat2011effects}, and the configuration-interaction method
\cite{johnson2013factorization}. Quantum gases are important topics in
statistical mechanics
\cite{firat2018characterization,dai2009exactly,firat2009universality,karabetoglu2017thermosize}%
. The energy spectrum of gas systems are studied, such as interacting as Bose
gases \cite{lewin2015bogoliubov}.

In section \ref{Clustercanonical}, we calculate the cluster expansion for the
canonical partition function. In section \ref{Clustercountingfunction}, we
calculate the cluster expansion for the spectral counting function. In section
\ref{Energyeigenvalue}, we construct an equation for energy eigenvalues. In
section \ref{Soft-sphere}, we calculate the energy spectrum of an interacting
gas with soft-sphere potentials. Conclusions are given in section
\ref{Conclusion}.

\section{Cluster expansion of energy spectrum for interacting many-body
system: scheme \label{scheme}}

The main aim of this paper is to convert the cluster expansion method in
statistical mechanics into a method for solving the energy eigenvalue of
interacting gases. By this method, the perturbation series of the energy
eigenvalue of an interacting gas can be expressed in terms of the cluster
expansion coefficient, as long as the interparticle potential is given.

\subsection{Equation of energy spectrum of interacting many-body system
\label{ZN}}

Calculating the energy spectrum from the canonical partition function follows
two steps.

\textit{Calculating spectral counting function from canonical partition
function. }Both the partition function and the counting function are spectral
functions, i.e., functions defined by eigenvalues. For the Hamiltonian whose
eigenvalue spectrum is $\left\{  E_{n}\right\}  $, the canonical partition
function is defined as

\begin{tcolorbox}[boxrule=0pt,
  boxsep=0pt,
  colback={lightgray1},
  enhanced jigsaw,
  borderline west={3pt}{0pt}{lightgray2},
  sharp corners,
  before skip=10pt,
  after skip=10pt,
breakable,]
\begin{equation}
Z\left(  \beta\right)  =\sum_{n}e^{-\beta E_{n}};
\end{equation}
\end{tcolorbox}
\noindent the spectral counting function $\Omega\left(  E\right)  $ is defined as the
number of eigenvalues less than $E$:

\begin{tcolorbox}[boxrule=0pt,
  boxsep=0pt,
  colback={lightgray1},
  enhanced jigsaw,
  borderline west={3pt}{0pt}{lightgray2},
  sharp corners,
  before skip=10pt,
  after skip=10pt,
breakable,]
\begin{equation}
\Omega\left(  E\right)  =\sum_{n}\theta\left(  E-E_{n}\right)  .
\label{NLamda}%
\end{equation}
\end{tcolorbox}
\noindent In Refs. \cite{dai2009number,dai2010approach,zhou2018calculating}, we prove a
relation between the spectral counting function and the canonical partition
function:%
\begin{tcolorbox}[boxrule=0pt,
  boxsep=0pt,
  colback={lightgray1},
  enhanced jigsaw,
  borderline west={3pt}{0pt}{lightgray2},
  sharp corners,
  before skip=10pt,
  after skip=10pt,
breakable,]
\begin{equation}
\Omega\left(  E\right)  =\mathcal{L}^{-1}\left[  \frac{Z\left(  \beta\right)
}{\beta}\right]  +c,\text{ \ \ \ }\left\{
\begin{array}
[c]{c}%
c=0,\text{ }E\neq E_{n},\\
c=-\frac{1}{2},\text{ }E=E_{n},
\end{array}
\right.  \label{NEIL}%
\end{equation}
\end{tcolorbox}
\noindent where $\mathcal{L}^{-1}$ denotes the inverse Laplace transform. By this
relation, we can calculate the spectral counting function from the canonical
partition function.

\textit{Calculating energy spectrum from spectral counting function.} The
$n$-th eigenvalue $E_{n}$ can be solved from the equation
\cite{courant2008methods,dai2009number}:%
\begin{tcolorbox}[boxrule=0pt,
  boxsep=0pt,
  colback={lightgray1},
  enhanced jigsaw,
  borderline west={3pt}{0pt}{lightgray2},
  sharp corners,
  before skip=10pt,
  after skip=10pt,
breakable,]

\begin{equation}
\Omega\left(  E_{n}\right)  =n. \label{NEn}%
\end{equation}
\end{tcolorbox}
\noindent As long as the spectral counting function is known, the energy spectrum can be solved.

\subsection{Cluster expansion of canonical partition function}

The cluster expansion method works in grand canonical ensembles. In order to
use the method of calculating the eigenvalue described in section \ref{ZN}, we
need canonical partition functions rather than grand canonical partition
functions. This requires us to construct cluster expansions for canonical
partition functions. Based on the mathematical result of the integer partition
function and the symmetric function
\cite{andrews1998theory,andrews2004integer,hardy1999ramanujan,zhou2018statistical}%
, we suggest a method which allows us to calculate the cluster expansion of
canonical partition functions from the cluster expansion of grand canonical
partition functions \cite{zhou2018canonical}.

In Ref. \cite{zhou2018canonical}, we give an expression of canonical partition
functions expressed by quantum and classical cluster expansion coefficients
for $N$-particle interacting gases:%
\begin{tcolorbox}[boxrule=0pt,
  boxsep=0pt,
  colback={lightgray1},
  enhanced jigsaw,
  borderline west={3pt}{0pt}{lightgray2},
  sharp corners,
  before skip=10pt,
  after skip=10pt,
breakable,]

\begin{equation}
Z\left(  \beta,N\right)  =\frac{1}{N!}B_{N}\left(  \Gamma_{1},\cdots
,\Gamma_{N}\right)  . \label{ZbetaBN}%
\end{equation}
\end{tcolorbox}
\noindent Here $B_{N}\left(  \Gamma_{1},\cdots,\Gamma_{N}\right)  $ is the Bell
polynomial \cite{zhou2018canonical}, where $\Gamma_{l}=\frac{l!V}{\lambda^{3}%
}b_{l}$ with $b_{l}$ the cluster-expansion coefficient
\cite{pathria2011statistical}, $\lambda=h\sqrt{\frac{\beta}{2\pi m}}$ the
thermal wavelength, and $V$ the volume. It should be noted that for classical
gases, $Z\left(  \beta,N\right)  =B_{N}\left(  \Gamma_{1},\cdots,\Gamma
_{N}\right)  $, while for identical-particle gases, $Z\left(  \beta,N\right)
=\frac{1}{N!}B_{N}\left(  \Gamma_{1},\cdots,\Gamma_{N}\right)  $.

\section{Cluster expansion of canonical partition function
\label{Clustercanonical}}

In this section, we calculate the cluster expansion of canonical partition functions.

The Bell polynomial in the canonical partition function (\ref{ZbetaBN}) can be
expressed as \cite{zhou2018canonical}%
\begin{equation}
B_{N}\left(  \Gamma_{1},\cdots,\Gamma_{N}\right)  =\det%
\begin{bmatrix}
\Gamma_{1} & C_{N-1}^{1}\Gamma_{2} & C_{N-1}^{2}\Gamma_{3} & \cdots & \cdots &
\Gamma_{N}\\
-1 & \Gamma_{1} & C_{N-2}^{1}\Gamma_{2} & \cdots & \cdots & \Gamma_{N-1}\\
0 & -1 & \Gamma_{1} & \cdots & \cdots & \Gamma_{N-2}\\
0 & 0 & -1 & \cdots & \cdots & \Gamma_{N-3}\\
\vdots & \vdots & \vdots & \ddots & \ddots & \vdots\\
0 & 0 & 0 & \cdots & -1 & \Gamma_{1}%
\end{bmatrix}
, \label{BellP}%
\end{equation}
where $C_{p}^{q}=\frac{p!}{q!\left(  p-q\right)  !}$. Writing the canonical
partition function (\ref{ZbetaBN}) as
\begin{equation}
Z\left(  \beta,N\right)  =\frac{1}{N!}\left(  \frac{V}{\lambda^{3}}\right)
^{N}\det\left(  1-A\right)  \label{ZbN}%
\end{equation}
with
\begin{equation}
A=%
\begin{bmatrix}
0 & -2b_{2} & -3b_{3} & \cdots & \cdots & -Nb_{N}\\
\left(  N-1\right)  \frac{\lambda^{3}}{V} & 0 & -2b_{2} & \cdots & \cdots &
-\left(  N-1\right)  b_{N-1}\\
0 & \left(  N-2\right)  \frac{\lambda^{3}}{V} & 0 & \cdots & \cdots & -\left(
N-2\right)  b_{N-2}\\
0 & 0 & \left(  N-3\right)  \frac{\lambda^{3}}{V} & \cdots & \cdots & -\left(
N-3\right)  b_{N-3}\\
\vdots & \vdots & \vdots & \ddots & \ddots & \vdots\\
0 & 0 & 0 & \cdots & 1\frac{\lambda^{3}}{V} & 0
\end{bmatrix}
. \label{A}%
\end{equation}
Note that here $b_{1}=1$.

We rewrite the determinant in Eq. (\ref{ZbN}) as%
\begin{align}
\det\left(  1-A\right)   &  =\exp\left(  \ln\left[  \det\left(  1-A\right)
\right]  \right) \nonumber\\
&  =\exp\left(  \operatorname{tr}\left[  \ln\left(  1-A\right)  \right]
\right)
\end{align}
and for weak interparticle interactions we expand the determinant:%
\begin{align}
\det\left(  1-A\right)   &  =\exp\left(  -\sum_{n=1}^{\infty}\frac
{\operatorname{tr}A^{n}}{n}\right) \nonumber\\
&  =1-\operatorname{tr}A-\frac{1}{2}\operatorname{tr}A^{2}-\frac{1}%
{3}\operatorname{tr}A^{3}+\cdots. \label{DetI-A}%
\end{align}
It should be noted that since according to Eq. (\ref{A}) $\operatorname{tr}%
A=0$, there are no cross terms in the first three orders of the expansion, and
the cross term will appear in higher-order terms. In this paper, we only
consider the first three-order contributions.

From Eq. (\ref{A}) we obtain%
\begin{align}
\operatorname{tr}A  &  =0,\nonumber\\
\operatorname{tr}A^{2}  &  =-2N\left(  N-1\right)  \frac{\lambda^{3}}{V}%
b_{2},\nonumber\\
\operatorname{tr}A^{3}  &  =-3N\left(  N-1\right)  \left(  N-2\right)  \left(
\frac{\lambda^{3}}{V}\right)  ^{2}b_{3},\nonumber\\
&  \cdots.
\end{align}
The canonical partition function then reads%
\begin{equation}
Z\left(  \beta,N\right)  =\frac{1}{N!}\left(  \frac{V}{\lambda^{3}}\right)
^{N}+\frac{1}{\left(  N-2\right)  !}\left(  \frac{V}{\lambda^{3}}\right)
^{N-1}b_{2}+\frac{1}{\left(  N-3\right)  !}\left(  \frac{V}{\lambda^{3}%
}\right)  ^{N-2}b_{3}+\cdots. \label{Z}%
\end{equation}
This is the canonical partition function of an $N$ particle interacting gas
expressed by the cluster expansion coefficient $b_{l}$.

The cluster expansion coefficient $b_{l}$ here is obtained in the grand
canonical ensemble, which can be expressed by the two-particle function
\begin{equation}
f_{ij}\equiv e^{-\beta u_{ij}}-1 \label{fij}%
\end{equation}
with $u_{ij}$ the two-particle interaction \cite{pathria2011statistical}:%
\begin{align}
b_{2}  &  =\frac{1}{2!\lambda^{3}V}\int f_{12}d^{3}r_{1}d^{3}r_{2},\nonumber\\
b_{3}  &  =\frac{1}{3!\lambda^{6}V}\int\left(  f_{12}f_{13}+f_{12}%
f_{23}+f_{13}f_{23}+f_{12}f_{13}f_{23}\right)  d^{3}r_{1}d^{3}r_{2}d^{3}%
r_{3},\nonumber\\
&  \cdots. \label{b2b3}%
\end{align}
Substituting Eq. (\ref{b2b3}) into Eq. (\ref{Z}) gives%
\begin{tcolorbox}[boxrule=0pt,
  boxsep=0pt,
  colback={lightgray1},
  enhanced jigsaw,
  borderline west={3pt}{0pt}{lightgray2},
  sharp corners,
  before skip=10pt,
  after skip=10pt,
breakable,]
\begin{align}
Z\left(  \beta,N\right)   &  =\frac{V^{N}}{N!}\frac{\left(  2\pi m\right)
^{\frac{3}{2}N}}{h^{3N}\beta^{\frac{3}{2}N}}+\frac{V^{N-2}}{2!\left(
N-2\right)  !}\frac{\left(  2\pi m\right)  ^{\frac{3}{2}N}}{h^{3N}\beta
^{\frac{3}{2}N}}\int f_{12}d^{3}r_{1}d^{3}r_{2}\nonumber\\
&  +\frac{V^{N-3}}{3!\left(  N-3\right)  !}\frac{\left(  2\pi m\right)
^{\frac{3}{2}N}}{h^{3N}\beta^{\frac{3}{2}N}}\int\left(  f_{12}f_{13}%
+f_{12}f_{23}+f_{13}f_{23}+f_{12}f_{13}f_{23}\right)  d^{3}r_{1}d^{3}%
r_{2}d^{3}r_{3}+\cdots. \label{ZbetaN2}%
\end{align}
\end{tcolorbox}

\section{Cluster expansion of spectral counting function
\label{Clustercountingfunction}}

According to the scheme illustrated in section \ref{scheme}, we calculate the
spectral counting function $\Omega\left(  E\right)  $ using the relation
(\ref{NEIL}). Substituting the canonical partition function (\ref{ZbetaN2})
into Eq. (\ref{NEIL}), we have
\begin{align}
\Omega\left(  E\right)   &  =\frac{V^{N}}{N!}\frac{\left(  2\pi m\right)
^{\frac{3N}{2}}}{h^{3N}}\mathcal{L}^{-1}\left[  \beta^{-\frac{3}{2}N-1}\right]
\nonumber\\
&  +\frac{V^{N-2}}{2!\left(  N-2\right)  !}\frac{\left(  2\pi m\right)
^{\frac{3N}{2}}}{h^{3N}}\mathcal{L}^{-1}\left[  \beta^{-\frac{3}{2}N-1}\int
d^{3}r_{1}d^{3}r_{2}f_{12}\right] \nonumber\\
&  +\frac{V^{N-3}}{3!\left(  N-3\right)  !}\frac{\left(  2\pi m\right)
^{\frac{3N}{2}}}{h^{3N}}\mathcal{L}^{-1}\left[  \beta^{-\frac{3}{2}N-1}\int
d^{3}r_{1}d^{3}r_{2}d^{3}r_{3}\left(  f_{12}f_{13}+f_{12}f_{23}+f_{13}%
f_{23}+f_{12}f_{13}f_{23}\right)  \right] \nonumber\\
&  +\cdots. \label{NEexpansion}%
\end{align}

Next, we calculate the inverse Laplace transform in Eq. (\ref{NEexpansion}) in
virtue of the convolution theorem \cite{debnath2016integral},%
\begin{equation}
\mathcal{L}^{-1}\left[  f\left(  s\right)  g\left(  s\right)  \right]
=F\left(  t\right)  \ast G\left(  t\right)  =\int_{0}^{t}F\left(
t-\tau\right)  G\left(  \tau\right)  d\tau, \label{juanji}%
\end{equation}
where $F\left(  t\right)  =\mathcal{L}^{-1}\left[  f\left(  s\right)  \right]
$ and $G\left(  t\right)  =\mathcal{L}^{-1}\left[  g\left(  s\right)  \right]
$.

The inverse Laplace transforms in Eq. (\ref{NEexpansion}) are
\begin{equation}
\mathcal{L}^{-1}\left[  \beta^{-\frac{3}{2}N-1}\right]  =\frac{E^{\frac{3N}%
{2}}}{\Gamma\left(  \frac{3N}{2}+1\right)  }\theta\left(  E\right)  ,
\end{equation}

\begin{align}
&  \mathcal{L}^{-1}\left[  \beta^{-\frac{3}{2}N-1}\int d^{3}r_{1}d^{3}%
r_{2}f_{12}\right]  =\mathcal{L}^{-1}\left[  \beta^{-\frac{3}{2}N-1}\right]
\ast\mathcal{L}^{-1}\left[  \int d^{3}r_{1}d^{3}r_{2}f_{12}\right] \nonumber\\
&  =\int d^{3}r_{1}d^{3}r_{2}\left[  \frac{E^{\frac{3N}{2}}}{\Gamma\left(
\frac{3N}{2}+1\right)  }\theta\left(  E\right)  \right]  \ast\tilde{f}_{12},
\end{align}

\begin{align}
&  \mathcal{L}^{-1}\left[  \beta^{-\frac{3}{2}N-1}\int d^{3}r_{1}d^{3}%
r_{2}d^{3}r_{3}\left(  f_{12}f_{13}+f_{12}f_{23}+f_{13}f_{23}+f_{12}%
f_{13}f_{23}\right)  \right] \nonumber\\
&  =\int d^{3}r_{1}d^{3}r_{2}d^{3}r_{3}\left[  \frac{E^{\frac{3N}{2}}}%
{\Gamma\left(  \frac{3N}{2}+1\right)  }\theta\left(  E\right)  \right]
\ast\left(  \tilde{f}_{12}\ast\tilde{f}_{13}+\tilde{f}_{12}\ast\tilde{f}%
_{23}+\tilde{f}_{13}\ast\tilde{f}_{23}+\tilde{f}_{12}\ast\tilde{f}_{13}%
\ast\tilde{f}_{23}\right)  ,
\end{align}
where $\tilde{f}_{ij}$ is the inverse Laplace transform of $f_{ij}$,%
\begin{equation}
\tilde{f}_{ij}=\mathcal{L}^{-1}\left[  f_{ij}\right]
\end{equation}
and $\theta\left(  E\right)  $ is the Heaviside theta function.

The spectral counting function (\ref{NEexpansion}) then reads%
\begin{align}
\Omega\left(  E\right)   &  =\frac{V^{N}}{N!}\frac{\left(  2\pi m\right)
^{\frac{3N}{2}}}{h^{3N}}\frac{E^{\frac{3N}{2}}}{\Gamma\left(  \frac{3N}%
{2}+1\right)  }\theta\left(  E\right) \nonumber\\
&  +\frac{V^{N-2}}{2!\left(  N-2\right)  !}\frac{\left(  2\pi m\right)
^{\frac{3N}{2}}}{h^{3N}}\int d^{3}r_{1}d^{3}r_{2}\left\{  \left[
\frac{E^{\frac{3N}{2}}}{\Gamma\left(  \frac{3N}{2}+1\right)  }\theta\left(
E\right)  \right]  \ast\tilde{f}_{12}\right\} \nonumber\\
&  +\frac{V^{N-3}}{3!\left(  N-3\right)  !}\frac{\left(  2\pi m\right)
^{\frac{3N}{2}}}{h^{3N}}\int d^{3}r_{1}d^{3}r_{2}d^{3}r_{3}\left\{  \left[
\frac{E^{\frac{3N}{2}}}{\Gamma\left(  \frac{3N}{2}+1\right)  }\theta\left(
E\right)  \right]  \right. \nonumber\\
&  \left.  \ast\left(  \tilde{f}_{12}\ast\tilde{f}_{13}+\tilde{f}_{12}%
\ast\tilde{f}_{23}+\tilde{f}_{13}\ast\tilde{f}_{23}+\tilde{f}_{12}\ast
\tilde{f}_{13}\ast\tilde{f}_{23}\right)  \right\} \nonumber\\
&  +\cdots. \label{NEfij}%
\end{align}

For the calculation of the second term in Eq. (\ref{NEfij}), we consider the
inverse Laplace transform of Eq. (\ref{fij}) as
\begin{align}
\tilde{f}_{ij}  &  =\mathcal{L}^{-1}\left[  e^{-\beta u_{ij}}-1\right]
\nonumber\\
&  =\delta\left(  E-u_{ij}\right)  -\delta\left(  E\right)
\end{align}
and the convolution in Eq. (\ref{NEfij}) can be calculated by Eq.
(\ref{juanji}):%
\begin{align}
\left[  \frac{E^{\frac{3N}{2}}}{\Gamma\left(  \frac{3N}{2}+1\right)  }%
\theta\left(  E\right)  \right]  \ast\tilde{f}_{12}  &  =\frac{E^{\frac{3N}%
{2}}}{\Gamma\left(  \frac{3N}{2}+1\right)  }\ast\left[  \delta\left(
E-u_{12}\right)  -\delta\left(  E\right)  \right] \nonumber\\
&  =\int_{0}^{E}\frac{\left(  E-\varepsilon\right)  ^{\frac{3N}{2}}}%
{\Gamma\left(  \frac{3N}{2}+1\right)  }\delta\left(  \varepsilon
-u_{12}\right)  d\varepsilon-\int_{0}^{E}\frac{\left(  E-\varepsilon\right)
^{\frac{3N}{2}}}{\Gamma\left(  \frac{3N}{2}+1\right)  }\delta\left(
\varepsilon\right)  d\varepsilon\nonumber\\
&  =\frac{E^{\frac{3N}{2}}}{\Gamma\left(  \frac{3N}{2}+1\right)  }\left[
\left(  1-\frac{u_{12}}{E}\right)  ^{\frac{3N}{2}}\theta\left(  E-u_{12}%
\right)  -\theta\left(  E\right)  \right]  , \label{NEfij1}%
\end{align}
where
\begin{align}
\int_{0}^{E}\frac{\left(  E-\varepsilon\right)  ^{\frac{3N}{2}}}{\Gamma\left(
\frac{3N}{2}+1\right)  }\delta\left(  \varepsilon-u_{12}\right)  d\varepsilon
&  =\left.  \frac{\left(  E-u_{12}\right)  ^{\frac{3N}{2}}}{\Gamma\left(
\frac{3N}{2}+1\right)  }\right\vert _{0<u_{12}<E}=\frac{\left(  E-u_{12}%
\right)  ^{\frac{3N}{2}}}{\Gamma\left(  \frac{3N}{2}+1\right)  }\theta\left(
E-u_{12}\right)  ,\nonumber\\
\int_{0}^{E}\frac{\left(  E-\varepsilon\right)  ^{\frac{3N}{2}}}{\Gamma\left(
\frac{3N}{2}+1\right)  }\delta\left(  \varepsilon\right)  d\varepsilon &
=\frac{E^{\frac{3N}{2}}}{\Gamma\left(  \frac{3N}{2}+1\right)  }\theta\left(
E\right)
\end{align}
are used.

For the calculation of the third term in Eq. (\ref{NEfij}), we consider the
inverse Laplace transform of $\tilde{f}_{ij}\ast\tilde{f}_{kl}$ and $\tilde
{f}_{ij}\ast\tilde{f}_{kl}\ast\tilde{f}_{mn}$.\textit{ }A direct calculation
gives%
\begin{equation}
f_{12}f_{13}+f_{12}f_{23}+f_{13}f_{23}+f_{12}f_{13}f_{23}=e^{-\beta\left(
u_{12}+u_{13}+u_{23}\right)  }-e^{-\beta u_{12}}-e^{-\beta u_{13}}-e^{-\beta
u_{23}}+2.
\end{equation}
Then
\begin{align}
&  \tilde{f}_{12}\ast\tilde{f}_{13}+\tilde{f}_{12}\ast\tilde{f}_{23}+\tilde
{f}_{13}\ast\tilde{f}_{23}+\tilde{f}_{12}\ast\tilde{f}_{13}\ast\tilde{f}%
_{23}\nonumber\\
&  =\mathcal{L}^{-1}\left[  f_{12}f_{13}+f_{12}f_{23}+f_{13}f_{23}%
+f_{12}f_{13}f_{23}\right] \nonumber\\
&  =\delta\left[  E-\left(  u_{12}+u_{13}+u_{23}\right)  \right]
-\delta\left(  E-u_{12}\right)  -\delta\left(  E-u_{13}\right)  -\delta\left(
E-u_{23}\right)  +2\delta\left(  E\right)  .
\end{align}
Therefore%
\begin{align}
&  \left[  \frac{E^{\frac{3N}{2}}}{\Gamma\left(  \frac{3N}{2}+1\right)
}\theta\left(  E\right)  \right]  \ast\left[  \int d^{3}r_{1}d^{3}r_{2}%
d^{3}r_{3}\left(  \tilde{f}_{12}\ast\tilde{f}_{13}+\tilde{f}_{12}\ast\tilde
{f}_{23}+\tilde{f}_{13}\ast\tilde{f}_{23}+\tilde{f}_{12}\ast\tilde{f}_{13}%
\ast\tilde{f}_{23}\right)  \right] \nonumber\\
&  =\int d^{3}r_{1}d^{3}r_{2}d^{3}r_{3}\frac{E^{\frac{3N}{2}}}{\Gamma\left(
\frac{3N}{2}+1\right)  }\left\{  \left(  1-\frac{u_{12}+u_{13}+u_{23}}%
{E}\right)  ^{\frac{3N}{2}}\theta\left[  E-\left(  u_{12}+u_{13}%
+u_{23}\right)  \right]  \right. \nonumber\\
&  \left.  -\left(  1-\frac{u_{12}}{E}\right)  ^{\frac{3N}{2}}\theta\left(
E-u_{12}\right)  -\left(  1-\frac{u_{13}}{E}\right)  ^{\frac{3N}{2}}%
\theta\left(  E-u_{13}\right)  -\left(  1-\frac{u_{23}}{E}\right)  ^{\frac
{3N}{2}}\theta\left(  E-u_{23}\right)  +2\theta\left(  E\right)  \right\}
\nonumber\\
&  =\frac{E^{\frac{3N}{2}}}{\Gamma\left(  \frac{3N}{2}+1\right)  }\left\{
\int d^{3}r_{1}d^{3}r_{2}d^{3}r_{3}\left(  1-\frac{u_{12}+u_{13}+u_{23}}%
{E}\right)  ^{\frac{3N}{2}}\theta\left[  E-\left(  u_{12}+u_{13}%
+u_{23}\right)  \right]  \right. \nonumber\\
&  -V\int d^{3}r_{1}d^{3}r_{2}\left(  1-\frac{u_{12}}{E}\right)  ^{\frac
{3N}{2}}\theta\left(  E-u_{12}\right)  -V\int d^{3}r_{1}d^{3}r_{3}\left(
1-\frac{u_{13}}{E}\right)  ^{\frac{3N}{2}}\theta\left(  E-u_{13}\right)
\nonumber\\
&  \left.  -V\int d^{3}r_{2}d^{3}r_{3}\left(  1-\frac{u_{23}}{E}\right)
^{\frac{3N}{2}}\theta\left(  E-u_{23}\right)  +2\theta\left(  E\right)
V^{3}\right\}  .
\end{align}

Then the spectral counting function (\ref{NEfij}) becomes%
\begin{align}
\Omega\left(  E\right)   &  =\frac{V^{N}}{N!}\frac{\left(  2\pi m\right)
^{\frac{3N}{2}}}{h^{3N}}\frac{E^{\frac{3N}{2}}}{\Gamma\left(  \frac{3N}%
{2}+1\right)  }\theta\left(  E\right)  +\frac{V^{N}}{2!\left(  N-2\right)
!}\frac{\left(  2\pi m\right)  ^{\frac{3N}{2}}}{h^{3N}}\frac{E^{\frac{3N}{2}}%
}{\Gamma\left(  \frac{3N}{2}+1\right)  }\nonumber\\
&  \times\left[  \frac{1}{V^{2}}\int d^{3}r_{1}d^{3}r_{2}\left(
1-\frac{u_{12}}{E}\right)  ^{\frac{3N}{2}}\theta\left(  E-u_{12}\right)
-\theta\left(  E\right)  \right]  +\frac{V^{N}}{3!\left(  N-3\right)  !}%
\frac{\left(  2\pi m\right)  ^{\frac{3N}{2}}}{h^{3N}}\frac{E^{\frac{3N}{2}}%
}{\Gamma\left(  \frac{3N}{2}+1\right)  }\nonumber\\
&  \times\left[  \frac{1}{V^{3}}\int d^{3}r_{1}d^{3}r_{2}d^{3}r_{3}\left(
1-\frac{u_{12}+u_{13}+u_{23}}{E}\right)  ^{\frac{3N}{2}}\theta\left(
E-\left(  u_{12}+u_{13}+u_{23}\right)  \right)  \right. \nonumber\\
&  -\frac{1}{V^{2}}\int d^{3}r_{1}d^{3}r_{2}\left(  1-\frac{u_{12}}{E}\right)
^{\frac{3N}{2}}\theta\left(  E-u_{12}\right)  -\frac{1}{V^{2}}\int d^{3}%
r_{1}d^{3}r_{3}\left(  1-\frac{u_{13}}{E}\right)  ^{\frac{3N}{2}}\theta\left(
E-u_{13}\right) \nonumber\\
&  \left.  -\frac{1}{V^{2}}\int d^{3}r_{2}d^{3}r_{3}\left(  1-\frac{u_{23}}%
{E}\right)  ^{\frac{3N}{2}}\theta\left(  E-u_{23}\right)  +2\theta\left(
E\right)  \right]  +\cdots. \label{NE2}%
\end{align}

Expanding (\ref{NE2}) by use of $\left(  1-\frac{a}{E}\right)  ^{\frac{3N}{2}%
}\simeq1-\frac{3N}{2}\frac{a}{E}+\frac{3N\left(  3N-2\right)  }{8}\frac{a^{2}%
}{E^{2}}+\cdots$ gives%
\begin{tcolorbox}[boxrule=0pt,
  boxsep=0pt,
  colback={lightgray1},
  enhanced jigsaw,
  borderline west={3pt}{0pt}{lightgray2},
  sharp corners,
  before skip=10pt,
  after skip=10pt,
breakable,]
\begin{align}
\Omega\left(  E\right)   &  =\sum_{k=0}^{\left[  \frac{3N}{2}\right]  }\left(
-1\right)  ^{k}\frac{c_{k}}{\Gamma\left(  1+\frac{3N}{2}-k\right)  }%
E^{\frac{3N}{2}-k}\nonumber\\
&  =\frac{c_{0}}{\Gamma\left(  \frac{3N}{2}+1\right)  }E^{\frac{3N}{2}}%
-\frac{c_{1}}{\Gamma\left(  \frac{3N}{2}\right)  }E^{\frac{3N}{2}-1}%
+\frac{c_{2}}{\Gamma\left(  \frac{3N}{2}-1\right)  }E^{\frac{3N}{2}-2}%
-\frac{c_{3}}{\Gamma\left(  \frac{3N}{2}-2\right)  }E^{\frac{3N}{2}-3}+\cdots,
\label{NEexpansion2}%
\end{align}
\end{tcolorbox}
\noindent where
\begin{align}
c_{0}  &  =\frac{V^{N}}{N!}\frac{\left(  2\pi m\right)  ^{\frac{3N}{2}}%
}{h^{3N}}\left\{  \frac{\left(  N-2\right)  \left(  N-3\right)  \left(
2N+1\right)  }{6}\theta\left(  E\right)  \right. \nonumber\\
&  -\frac{N\left(  N-1\right)  \left(  N-5\right)  }{3!V^{2}}\int d^{3}%
r_{1}d^{3}r_{2}\theta\left(  E-u_{12}\right)  -\frac{N\left(  N-1\right)
\left(  N-2\right)  }{3!V^{2}}\int d^{3}r_{1}d^{3}r_{3}\theta\left(
E-u_{13}\right) \nonumber\\
&  \left.  -\frac{N\left(  N-1\right)  \left(  N-2\right)  }{3!V^{2}}\int
d^{3}r_{2}d^{3}r_{3}\theta\left(  E-u_{23}\right)  \right. \nonumber\\
&  \left.  +\frac{N\left(  N-1\right)  \left(  N-2\right)  }{3!V^{3}}\int
d^{3}r_{1}d^{3}r_{2}d^{3}r_{3}\theta\left[  E-\left(  u_{12}+u_{13}%
+u_{23}\right)  \right]  \right\}  , \label{c0}%
\end{align}%
\begin{align}
c_{1}=  &  \frac{V^{N}}{2!\left(  N-2\right)  !}\frac{\left(  2\pi m\right)
^{\frac{3N}{2}}}{h^{3N}}\left\{  \frac{N-2}{3V^{3}}\int d^{3}r_{1}d^{3}%
r_{2}d^{3}r_{3}\left(  u_{12}+u_{13}+u_{23}\right)  \right. \nonumber\\
&  \times\left.  \theta\left[  E-\left(  u_{12}+u_{13}+u_{23}\right)  \right]
\right\}  -\frac{N-5}{3V^{2}}\int d^{3}r_{1}d^{3}r_{2}u_{12}\theta\left(
E-u_{12}\right) \nonumber\\
&  -\frac{N-2}{3V^{2}}\int d^{3}r_{1}d^{3}r_{3}u_{13}\theta\left(
E-u_{13}\right)  \left.  -\frac{N-2}{3V^{2}}\int d^{3}r_{2}d^{3}r_{3}%
u_{23}\theta\left(  E-u_{23}\right)  \right\}  , \label{c1}%
\end{align}%
\begin{align}
c_{2}=\frac{1}{2!}  &  \frac{V^{N}}{2!\left(  N-2\right)  !}\frac{\left(  2\pi
m\right)  ^{\frac{3N}{2}}}{h^{3N}}\left\{  \frac{N-2}{3V^{3}}\int d^{3}%
r_{1}d^{3}r_{2}d^{3}r_{3}\left(  u_{12}+u_{13}+u_{23}\right)  ^{2}\right.
\nonumber\\
&  \times\left.  \theta\left[  E-\left(  u_{12}+u_{13}+u_{23}\right)  \right]
\right\}  -\frac{N-5}{3V^{2}}\int d^{3}r_{1}d^{3}r_{2}u_{12}^{2}\theta\left(
E-u_{12}\right) \nonumber\\
&  -\frac{N-2}{3V^{2}}\int d^{3}r_{1}d^{3}r_{3}u_{13}^{2}\theta\left(
E-u_{13}\right)  \left.  -\frac{N-2}{3V^{2}}\int d^{3}r_{2}d^{3}r_{3}%
u_{23}^{2}\theta\left(  E-u_{23}\right)  \right\}  . \label{c2}%
\end{align}

\section{Energy eigenvalue \label{Energyeigenvalue}}

Now we solve the energy eigenvalue. By Eqs. (\ref{NEn}) and
(\ref{NEexpansion2}), we arrive at an equation of the energy eigenvalue
$E_{n}$:%
\begin{equation}
\sum_{k=0}^{\left[  \frac{3N}{2}\right]  }\left(  -1\right)  ^{k}\frac{c_{k}%
}{\Gamma\left(  1+\frac{3N}{2}-k\right)  }E_{n}^{\frac{3N}{2}-k}=n. \label{6}%
\end{equation}

We first consider the leading order contribution:%
\begin{equation}
\frac{c_{0}}{\Gamma\left(  \frac{3N}{2}+1\right)  }E_{n}^{\frac{3N}{2}}=n
\end{equation}
which gives
\begin{equation}
E_{n}=\frac{\Gamma^{\frac{2}{3N}}\left(  \frac{3N}{2}+1\right)  }{c_{0}%
^{\frac{2}{3N}}}n^{\frac{2}{3N}}.
\end{equation}
To solve the equation, we expand $E_{n}$ as \cite{dai2009number}%
\begin{equation}
E_{n}=\frac{\Gamma^{\frac{2}{3N}}\left(  \frac{3N}{2}+1\right)  }{c_{0}%
^{\frac{2}{3N}}}n^{\frac{2}{3N}}+\sum_{m=0}^{\infty}\alpha_{m+1}n^{-\frac
{2m}{3N}}.
\end{equation}
Then Eq. (\ref{6}) becomes%
\begin{equation}
\frac{c_{0}}{\Gamma\left(  \frac{3N}{2}+1\right)  }\left[  \frac{\Gamma
^{\frac{2}{3N}}\left(  \frac{3N}{2}+1\right)  }{c_{0}^{\frac{2}{3N}}}%
n^{\frac{2}{3N}}+\alpha_{1}\right]  ^{\frac{3N}{2}}-\frac{c_{1}}{\Gamma\left(
\frac{3N}{2}\right)  }\left[  \frac{\Gamma^{\frac{2}{3N}}\left(  \frac{3N}%
{2}+1\right)  }{c_{0}^{\frac{2}{3N}}}n^{\frac{2}{3N}}+\alpha_{1}\right]
^{\frac{3N}{2}-1}+\cdots=n. \label{eqEn}%
\end{equation}

By the Newton binomial theorem \cite{weisstein2002crc}%
\begin{equation}
\left(  x+y\right)  ^{\alpha}=\sum_{k=0}^{\infty}\binom{\alpha}{k}x^{\alpha
-k}y^{k},
\end{equation}
where $\alpha$ is a real number and $\binom{\alpha}{k}=\frac{\Gamma\left(
\alpha+1\right)  }{\Gamma\left(  \alpha-k+1\right)  \Gamma\left(  k+1\right)
}$, we have
\begin{equation}
\left[  \frac{\Gamma^{\frac{2}{3N}}\left(  \frac{3N}{2}+1\right)  }%
{c_{0}^{\frac{2}{3N}}}n^{\frac{2}{3N}}+\alpha_{1}n^{0}\right]  ^{\frac{3N}%
{2}-r}=\sum_{j=0}^{\frac{3N}{2}-r}\binom{\frac{3N}{2}-r}{j}\left[
\frac{\Gamma^{\frac{2}{3N}}\left(  \frac{3N}{2}+1\right)  }{c_{0}^{\frac
{2}{3N}}}n^{\frac{2}{3N}}\right]  ^{\frac{3N}{2}-r-j}\alpha_{1}^{j},\text{
}0\leqslant r\leqslant\frac{3N}{2}.
\end{equation}
Then Eq. (\ref{eqEn}) becomes
\begin{align}
&  \frac{c_{0}}{\Gamma\left(  \frac{3N}{2}+1\right)  }\left\{  \sum
_{j=0}^{\frac{3N}{2}}\binom{\frac{3N}{2}}{j}\left[  \frac{\Gamma^{\frac{2}%
{3N}}\left(  \frac{3N}{2}+1\right)  }{c_{0}^{\frac{2}{3N}}}n^{\frac{2}{3N}%
}\right]  ^{\frac{3N}{2}-j}\alpha_{1}^{j}\right\} \nonumber\\
&  -\frac{c_{1}}{\Gamma\left(  \frac{3N}{2}\right)  }\left\{  \sum
_{j=0}^{\frac{3N}{2}-1}\binom{\frac{3N}{2}-1}{j}\left[  \frac{\Gamma^{\frac
{2}{3N}}\left(  \frac{3N}{2}+1\right)  }{c_{0}^{\frac{2}{3N}}}n^{\frac{2}{3N}%
}\right]  ^{\frac{3N}{2}-1-j}\alpha_{1}^{j}\right\}  +\cdots=n.
\end{align}
Keeping the first- and second-order contributions, we have
\begin{align}
&  \frac{c_{0}}{\Gamma\left(  \frac{3N}{2}+1\right)  }\left\{  \frac
{\Gamma\left(  \frac{3N}{2}+1\right)  }{c_{0}}n+\binom{\frac{3N}{2}}{1}\left[
\frac{\Gamma^{\frac{2}{3N}}\left(  \frac{3N}{2}+1\right)  }{c_{0}^{\frac
{2}{3N}}}n^{\frac{2}{3N}}\right]  ^{\frac{3N}{2}-1}\alpha_{1}\right\}
\nonumber\\
&  -\frac{c_{1}}{\Gamma\left(  \frac{3N}{2}\right)  }\left[  \frac
{\Gamma^{\frac{2}{3N}}\left(  \frac{3N}{2}+1\right)  }{c_{0}^{\frac{2}{3N}}%
}n^{\frac{2}{3N}}\right]  ^{\frac{3N}{2}-1}=n.
\end{align}
Equaling the coefficients at each power of $n$, we have%
\begin{equation}
\alpha_{1}=\frac{2}{3N}\frac{\Gamma\left(  \frac{3N}{2}+1\right)  }%
{\Gamma\left(  \frac{3N}{2}\right)  }\frac{c_{1}}{c_{0}}.
\end{equation}
Substituting $\alpha_{1}$ into $E_{n}$, we obtain the energy eigenvalue,%
\begin{equation}
E_{n}=\frac{\Gamma^{\frac{2}{3N}}\left(  \frac{3N}{2}+1\right)  }{c_{0}%
^{\frac{2}{3N}}}n^{\frac{2}{3N}}+\frac{2}{3N}\frac{\Gamma\left(  \frac{3N}%
{2}+1\right)  }{\Gamma\left(  \frac{3N}{2}\right)  }\frac{c_{1}}{c_{0}}%
n^{0}+\cdots. \label{En1}%
\end{equation}

For higher orders, we use the multinomial theorem \cite{weisstein2002crc}%
\begin{equation}
\left(  x_{1}+x_{2}+\cdots+x_{t}\right)  ^{p}=\sum_{\substack{p_{i}%
\geqslant0\\p_{1}+p_{2}+\cdots+p_{t}=p}}\binom{p}{p_{1},p_{2},\cdots,p_{t}%
}\prod\limits_{1\leqslant i\leqslant t}x_{i}^{p_{i}},
\end{equation}
where $\binom{p}{p_{1},p_{2},\cdots,p_{t}}=\frac{\Gamma\left(  p+1\right)
}{\Gamma\left(  p_{1}+1\right)  \Gamma\left(  p_{2}+1\right)  \cdots
\Gamma\left(  p_{t}+1\right)  }$. Similar calculation gives%
\begin{equation}
a_{2}=\frac{3N-2}{6N}\frac{c_{1}^{2}}{c_{0}^{2-\frac{2}{3N}}}\frac
{\Gamma^{2-\frac{2}{3N}}\left(  \frac{3N}{2}+1\right)  }{\Gamma^{2}\left(
\frac{3N}{2}\right)  }-\frac{\Gamma^{1-\frac{2}{3N}}\left(  \frac{3N}%
{2}+1\right)  }{\Gamma\left(  \frac{3N}{2}-1\right)  }\frac{c_{2}}%
{c_{0}^{1-\frac{2}{3N}}}.
\end{equation}
Then the energy eigenvalue reads
\begin{tcolorbox}[boxrule=0pt,
  boxsep=0pt,
  colback={lightgray1},
  enhanced jigsaw,
  borderline west={3pt}{0pt}{lightgray2},
  sharp corners,
  before skip=10pt,
  after skip=10pt,
breakable,]
\begin{align}
E_{n}  &  =\frac{\Gamma^{\frac{2}{3N}}\left(  \frac{3N}{2}+1\right)  }%
{c_{0}^{\frac{2}{3N}}}n^{\frac{2}{3N}}+\frac{2}{3N}\frac{\Gamma\left(
\frac{3N}{2}+1\right)  }{\Gamma\left(  \frac{3N}{2}\right)  }\frac{c_{1}%
}{c_{0}}n^{0}\nonumber\\
&  +\left[  \frac{3N-2}{6N}\frac{c_{1}^{2}}{c_{0}^{2-\frac{2}{3N}}}%
\frac{\Gamma^{2-\frac{2}{3N}}\left(  \frac{3N}{2}+1\right)  }{\Gamma
^{2}\left(  \frac{3N}{2}\right)  }-\frac{\Gamma^{1-\frac{2}{3N}}\left(
\frac{3N}{2}+1\right)  }{\Gamma\left(  \frac{3N}{2}-1\right)  }\frac{c_{2}%
}{c_{0}^{1-\frac{2}{3N}}}\right]  n^{-\frac{2}{3N}}+\cdots. \label{En}%
\end{align}
\end{tcolorbox}

In an interacting gas, there are three typical length scales: the range of the
interparticle interaction, the mean particle distance, and the thermal
wavelength. The range of the interparticle interaction is determined by the
interparticle potential (e.g., the radius of the\ gaseous molecule of a
soft-sphere gas), the mean particle distance is determined by the density of
the gas, and the thermal wavelength is determined by the temperature. The
comparison between these three typical length scales de termines the
properties of the gas; for example, if the thermal wavelength is greater than
the mean particle distance, it is a quantum gas, and so on. 

The density that determines the distance between gaseous molecules will
influence the energy spectrum of a gas. This requires us to discuss the energy
spectrum case by case with different gas densities.

The temperature determines the thermal wavelength. For a given density, the
longer the thermal wavelength is, the larger the overlapping region of the
particle wave function and the stronger the quantum exchange interaction, and
vice versa.

For the cluster expansion method, the effect of the interparticle interaction
is reflected in the factor $f_{ij}\equiv e^{-u_{ij}/\left(  kT\right)  }-1$
defined by Eq. (\ref{fij}). It can be seen from $f_{ij}$ that the effect of
the interparticle interaction decreases at high temperatures and increases at
low temperatures. Although the expression of the eigenvalue (\ref{En}) does
not include the temperature, the temperature will influence the eigenvalue.
The energy spectrum of gas is different at different temperatures.

Concretely, the effect of temperature on the partition function leads to the
effect of temperature on the counting function and then leads to the effect of
temperature on the expansion coefficient $c_{k}$\ in the energy spectrum. The
coefficient $c_{k}$ ($k\neq0$) decreases with increasing temperature. The
greater $k$, the faster $c_{k}$ decreases with increasing temperature. In
other words, although the temperature does not appear in $c_{k}$, the effect
of temperature influences $c_{k}$ through its effect on $f_{ij}$. The
higher-order contribution in the eigenvalue decreases with the temperature
increase. Higher-order contributions will become important in low-temperature systems.

\section{Soft-sphere gas \label{Soft-sphere}}

The soft-sphere potential is
\begin{equation}
u\left(  r\right)  =\left\{
\begin{array}
[c]{c}%
0,\text{ \ \ }r>D\\
v_{0},\text{ \ \ }r\leqslant D
\end{array}
\right.  ,
\end{equation}
where $v_{0}>0$.

As discussed in section \ref{Energyeigenvalue}, the density influences the
eigenvalue of the gas.\ The eigenvalue of gases with different densities is
different. In the following, we discuss the energy spectrum in four density
conditions, from high densities to low densities: high-density,
medium-high-density, medium-low-density, and low-density cases.

\textbf{\color{gray} High-density case}: In the high-density case, $\left\vert
\mathbf{r}_{1}-\mathbf{r}_{2}\right\vert <D$, $\left\vert \mathbf{r}%
_{1}-\mathbf{r}_{3}\right\vert <D$, and $\left\vert \mathbf{r}_{2}%
-\mathbf{r}_{3}\right\vert <D$, i.e., $E>3v_{0}\times\frac{N\left(
N-1\right)  \left(  N-2\right)  }{3!}$, we have
\begin{align}
&  \int d^{3}r_{i}d^{3}r_{j}\theta\left(  E-u_{ij}\right)  =V^{2},\nonumber\\
&  \int d^{3}r_{i}d^{3}r_{j}u_{ij}\theta\left(  E-u_{ij}\right)  =V\frac{4\pi
D^{3}}{3}v_{0},\nonumber\\
&  \int d^{3}r_{1}d^{3}r_{2}d^{3}r_{3}\theta\left[  E-\left(  u_{12}%
+u_{13}+u_{23}\right)  \right]  =V\frac{5\pi^{2}D^{6}}{6},\nonumber\\
&  \int d^{3}r_{1}d^{3}r_{2}d^{3}r_{3}\left(  u_{12}+u_{13}+u_{23}\right)
\theta\left[  E-\left(  u_{12}+u_{13}+u_{23}\right)  \right]  =V\frac{5\pi
^{2}D^{6}}{6}3v_{0}.
\end{align}
Then by Eqs. (\ref{c0}) and (\ref{c1}), we have
\begin{align}
c_{0}  &  =\frac{V^{N}}{N!}\frac{\left(  2\pi m\right)  ^{\frac{3N}{2}}%
}{h^{3N}}\left[  \binom{N}{3}\frac{5\pi^{2}D^{6}}{6V^{2}}-\frac{\left(
N^{2}+2\right)  \left(  N-3\right)  }{6}\right]  ,\\
c_{1}  &  =\frac{V^{N}}{2!\left(  N-2\right)  !}\frac{\left(  2\pi m\right)
^{\frac{3N}{2}}}{h^{3N}}v_{0}\left[  \left(  N-2\right)  \frac{5\pi^{2}D^{6}%
}{6V^{2}}-\left(  N-3\right)  \frac{4\pi D^{3}}{3V}\right]  .
\end{align}
The energy eigenvalue of a soft-sphere gas, by Eq. (\ref{En1}), reads%
\begin{align}
E_{n}  &  =E_{n}^{\text{free}}\frac{\left(  N!\right)  ^{\frac{2}{3N}}%
}{\left[  \binom{N}{3}\frac{5\pi^{2}D^{6}}{6V^{2}}-\frac{\left(
N^{2}+2\right)  \left(  N-3\right)  }{6}\right]  ^{\frac{2}{3N}}}\nonumber\\
&  +\frac{N\left(  N-1\right)  }{2}v_{0}\frac{\left(  N-2\right)  \frac
{5\pi^{2}D^{6}}{6V^{2}}-\left(  N-3\right)  \frac{4\pi D^{3}}{3V}}{\binom
{N}{3}\frac{5\pi^{2}D^{6}}{6V^{2}}-\frac{\left(  N^{2}+2\right)  \left(
N-3\right)  }{6}}n^{0}+\cdots,
\end{align}
where $E_{n}^{\text{free}}$ is the energy of ideal gases:
\begin{equation}
E_{n}^{\text{free}}=\frac{h^{2}}{2\pi mV^{\frac{2}{3}}}\Gamma^{\frac{2}{3N}%
}\left(  \frac{3}{2}N+1\right)  n^{\frac{2}{3N}}.
\end{equation}

\textbf{\color{gray} Medium-high-density case}: In the Medium-high-density case,
$\left\vert \mathbf{r}_{1}-\mathbf{r}_{2}\right\vert <D$, $\left\vert
\mathbf{r}_{1}-\mathbf{r}_{3}\right\vert <D$, $\left\vert \mathbf{r}%
_{2}-\mathbf{r}_{3}\right\vert >D$ or $\left\vert \mathbf{r}_{1}%
-\mathbf{r}_{2}\right\vert <D$, $\left\vert \mathbf{r}_{1}-\mathbf{r}%
_{3}\right\vert >D$, and $\left\vert \mathbf{r}_{2}-\mathbf{r}_{3}\right\vert
<D$, i.e., $2v_{0}\times\frac{N\left(  N-1\right)  \left(  N-2\right)  }%
{3!}<E<3v_{0}\times\frac{N\left(  N-1\right)  \left(  N-2\right)  }{3!}$, we
have%
\begin{align}
&  \int d^{3}r_{i}d^{3}r_{j}\theta\left(  E-u_{ij}\right)  =V^{2},\nonumber\\
&  \int d^{3}r_{i}d^{3}r_{j}u_{ij}\theta\left(  E-u_{ij}\right)  =V\frac{4\pi
D^{3}}{3}v_{0},\nonumber\\
&  \int d^{3}r_{1}d^{3}r_{2}d^{3}r_{3}\theta\left[  E-\left(  u_{12}%
+u_{13}+u_{23}\right)  \right]  =\frac{17V}{32}\left(  \frac{4\pi D^{3}}%
{3}\right)  ^{2},\nonumber\\
&  \int d^{3}r_{1}d^{3}r_{2}d^{3}r_{3}\left(  u_{12}+u_{13}+u_{23}\right)
\theta\left[  E-\left(  u_{12}+u_{13}+u_{23}\right)  \right]  =2v_{0}%
\frac{17V}{32}\left(  \frac{4\pi D^{3}}{3}\right)  ^{2}.
\end{align}
Then by Eqs. (\ref{c0}) and (\ref{c1}), we have%
\begin{align}
c_{0}  &  =\frac{V^{N}}{N!}\frac{\left(  2\pi m\right)  ^{\frac{3N}{2}}%
}{h^{3N}}\left[  \frac{17}{32}\binom{N}{3}\frac{\left(  \frac{4\pi D^{3}}%
{3}\right)  ^{2}}{V^{2}}-\frac{\left(  N^{2}+2\right)  \left(  N-3\right)
}{6}\right]  ,\\
c_{1}  &  =\frac{V^{N}}{2!\left(  N-2\right)  !}\frac{\left(  2\pi m\right)
^{\frac{3N}{2}}}{h^{3N}}\frac{4\pi D^{3}}{3V}v_{0}\left[  \frac{17\left(
N-2\right)  }{48}\frac{4\pi D^{3}}{3V}-\left(  N-3\right)  \right]  .
\end{align}
The energy eigenvalue of a soft-sphere gas, by Eq. (\ref{En1}), reads%
\begin{align}
E_{n}  &  =E_{n}^{\text{free}}\frac{\left(  N!\right)  ^{\frac{2}{3N}}%
}{\left[  \frac{17}{32}\binom{N}{3}\frac{\left(  \frac{4\pi D^{3}}{3}\right)
^{2}}{V^{2}}-\frac{\left(  N^{2}+2\right)  \left(  N-3\right)  }{6}\right]
^{\frac{2}{3N}}}\nonumber\\
&  +\frac{N\left(  N-1\right)  }{2}\frac{4\pi D^{3}}{3V}v_{0}\frac
{\frac{17\left(  N-2\right)  }{48}\frac{4\pi D^{3}}{3V}-\left(  N-3\right)
}{\frac{17}{32}\binom{N}{3}\frac{\left(  \frac{4\pi D^{3}}{3}\right)  ^{2}%
}{V^{2}}-\frac{\left(  N^{2}+2\right)  \left(  N-3\right)  }{6}}n^{0}+\cdots.
\end{align}

\textbf{\color{gray} Medium-low-density case}: In the Medium-low-density case, $\left\vert
\mathbf{r}_{1}-\mathbf{r}_{2}\right\vert <D$, $\left\vert \mathbf{r}%
_{1}-\mathbf{r}_{3}\right\vert >D$, $\left\vert \mathbf{r}_{2}-\mathbf{r}%
_{3}\right\vert >D$, i.e., $v_{0}\times\frac{N\left(  N-1\right)  \left(
N-2\right)  }{3!}<E<2v_{0}\times\frac{N\left(  N-1\right)  \left(  N-2\right)
}{3!}$, we have%
\begin{align}
&  \int d^{3}r_{i}d^{3}r_{j}\theta\left(  E-u_{ij}\right)  =V^{2},\nonumber\\
&  \int d^{3}r_{i}d^{3}r_{j}u_{ij}\theta\left(  E-u_{ij}\right)  =V\frac{4\pi
D^{3}}{3}v_{0},\nonumber\\
&  \int d^{3}r_{1}d^{3}r_{2}d^{3}r_{3}\theta\left[  E-\left(  u_{12}%
+u_{13}+u_{23}\right)  \right]  =V^{2}\frac{4\pi D^{3}}{3}\left(
1-\frac{49\pi D^{3}}{24V}\right)  ,\nonumber\\
&  \int d^{3}r_{1}d^{3}r_{2}d^{3}r_{3}\left(  u_{12}+u_{13}+u_{23}\right)
\theta\left[  E-\left(  u_{12}+u_{13}+u_{23}\right)  \right]  =V^{2}\frac{4\pi
D^{3}}{3}v_{0}\left(  1-\frac{49\pi D^{3}}{24V}\right)  .
\end{align}
Then by Eqs. (\ref{c0}) and (\ref{c1}), we have%
\begin{align}
c_{0}  &  =\frac{V^{N}}{N!}\frac{\left(  2\pi m\right)  ^{\frac{3N}{2}}%
}{h^{3N}}\left[  \binom{N}{3}\frac{4\pi D^{3}}{3V}\left(  1-\frac{49\pi D^{3}%
}{24V}\right)  -\frac{\left(  N^{2}+2\right)  \left(  N-3\right)  }{6}\right]
,\\
c_{1}  &  =-\frac{V^{N}}{2!\left(  N-2\right)  !}\frac{\left(  2\pi m\right)
^{\frac{3N}{2}}}{h^{3N}}\frac{4\pi D^{3}}{3V}\frac{v_{0}}{3}\left[  \left(
2N-7\right)  +\left(  N-2\right)  \frac{49\pi D^{3}}{24V}\right]  .
\end{align}
The energy eigenvalue of a soft-sphere gas, by Eq. (\ref{En1}), reads%
\begin{align}
E_{n}  &  =E_{n}^{\text{free}}\frac{\left(  N!\right)  ^{\frac{2}{3N}}%
}{\left[  \binom{N}{3}\frac{4\pi D^{3}}{3V}\left(  1-\frac{49\pi D^{3}}%
{24V}\right)  -\frac{\left(  N^{2}+2\right)  \left(  N-3\right)  }{6}\right]
^{\frac{2}{3N}}}\nonumber\\
&  -\frac{N\left(  N-1\right)  }{6}v_{0}\frac{4\pi D^{3}}{3V}\frac{\left(
2N-7\right)  +\left(  N-2\right)  \frac{49\pi D^{3}}{24V}}{\binom{N}{3}%
\frac{4\pi D^{3}}{3V}\left(  1-\frac{49\pi D^{3}}{24V}\right)  -\frac{\left(
N^{2}+2\right)  \left(  N-3\right)  }{6}}n^{0}+\cdots,v_{0}<E<2v_{0}.
\end{align}

\textbf{\color{gray} Low-density case}: In the low-density case, $\left\vert \mathbf{r}%
_{1}-\mathbf{r}_{2}\right\vert >D$, $\left\vert \mathbf{r}_{1}-\mathbf{r}%
_{3}\right\vert >D$, $\left\vert \mathbf{r}_{2}-\mathbf{r}_{3}\right\vert >D$,
i.e., $0<E<v_{0}\times\frac{N\left(  N-1\right)  \left(  N-2\right)  }{3!}$,
we have%
\begin{align}
&  \int d^{3}r_{i}d^{3}r_{j}\theta\left(  E-u_{ij}\right)  =V\left(
V-\frac{4\pi D^{3}}{3}\right)  ,\nonumber\\
&  \int d^{3}r_{i}d^{3}r_{j}u_{ij}\theta\left(  E-u_{ij}\right)
=0,\nonumber\\
&  \int d^{3}r_{1}d^{3}r_{2}d^{3}r_{3}\theta\left[  E-\left(  u_{12}%
+u_{13}+u_{23}\right)  \right]  =V^{3}\left(  1-\frac{4\pi D^{3}}{3V}\right)
,\nonumber\\
&  \int d^{3}r_{1}d^{3}r_{2}d^{3}r_{3}\left(  u_{12}+u_{13}+u_{23}\right)
\theta\left[  E-\left(  u_{12}+u_{13}+u_{23}\right)  \right]  =0.
\end{align}
Then by Eqs. (\ref{c0}) and (\ref{c1}), we have%
\begin{align}
c_{0}  &  =\frac{V^{N}}{N!}\frac{\left(  2\pi m\right)  ^{\frac{3N}{2}}%
}{h^{3N}}\left[  1+\binom{N}{2}\frac{2N-7}{3}\frac{4\pi D^{3}}{3V}\right]  ,\\
c_{1}  &  =0.
\end{align}
The energy eigenvalue of a soft-sphere gas, by Eq. (\ref{En1}), reads%
\begin{equation}
E_{n}=E_{n}^{\text{free}}\left(  N!\right)  ^{\frac{2}{3N}}\left[  \frac
{1}{1+\binom{N}{2}\frac{2N-7}{3}\frac{4\pi D^{3}}{3V}}\right]  ^{\frac{2}{3N}%
}.
\end{equation}

\section{Conclusions and outlook \label{Conclusion}}

In this paper, we provide a formula for the energy eigenvalue of interacting
gases, by converting the cluster method in statistical mechanics into a method
for calculating energy eigenvalues.

The steps to solve the energy eigenvalue spectrum are as follows. (1)
Calculate the cluster expansion of canonical partition functions from the
cluster expansion of grand partition functions. (2) Calculate the cluster
expansion of the spectral counting function from the cluster expansion of the
canonical partition function. (3) Calculate energy eigenvalues from spectral
counting functions.

Using the method, we calculate the energy eigenvalue of an interacting gas
with soft-sphere potentials.

The method used in this paper is to convert various statistical mechanical
methods into methods of finding the energy spectrum of many-body systems. In
principle, all kinds of statistical mechanics methods for solving partition
functions, grand partition functions, and thermodynamic quantities can be
converted into methods for solving energy spectra.

In this paper, we convert the cluster expansion method in statistical
mechanics into a method to calculate the energy spectrum of interacting gases.
The cluster expansion method applies to the week degenerate gases, so the
resulting energy spectrum is also for week degenerate cases. Although the most
direct way to obtain the strong degenerate result is to convert some kind of
statistical mechanical method for dealing with strong degenerate gases into a
method for solving energy spectra, there is still a possibility of extending
the cluster expansion method which originally applies only to weak degenerate
gases into a method for dealing with strong degenerate gases. This method is
based on the Pad\'{e} approximant. In Ref. \cite{tian2021pade}, we suggest a
method that converts the virtual expansion method for high-temperature and
low-density gases into a method applying to low-temperature and high-density
gases. After the Pad\'{e} approximant treatment, the virtual expansion method
can be used to consider the BEC phase transition and calculate the
low-temperature properties of Fermi gases. Similarly, the Pad\'{e}
approximation method suggested in Ref. \cite{tian2021pade} can also be used to
extend the cluster expansion method to strong degenerate cases. Similar to the
virial expansion, the cluster expansion also gives a truncated power series
approximation, which is essentially a polynomial approximation. From the power
series given by the cluster expansion, we can construct a rational function
approximation by the Pad\'{e} approximant. As can be seen from the experience
in Ref. \cite{tian2021pade}, this treatment will extend the method for weak
degenerate to a method for strong degenerate.

The counting function, Green function, heat kernel \cite{vassilevich2003heat},
and even scattering phase shift \cite{pang2012relation,li2015heat} of a given
operator $H$ are all spectral functions, defined by the eigenvalues of the
operator. This work is based on the relation between counting functions and
partition functions. The partition function is the trace of the local heat
kernel which is the Green function of the initial-value problem of the
heat-type equation defined by the operator $H$ \cite{vassilevich2003heat}.
There is also a trace formula for the operator $H$, which corresponds to its
Green function
\cite{stockmann1999quantum,cvitanovic2015chaos,gutzwiller2013chaos}. In future
works, we will consider calculating eigenvalues by trace formula from the
Green function.

\acknowledgments

We are very indebted to Dr G. Zeitrauman for his encouragement. This work is supported in part by Special Funds for theoretical physics Research Program of the NSFC under Grant No.
11947124, and NSFC under Grant Nos. 11575125 and 11675119.

\nolinenumbers
%%%%%%%%%%%%%%%%%%%%%%%%%%%%%%%%%%%%%%%%%%

%%%%%%%%%%%%%%%%%%%%%%%%%%%%%%%%%%%%%%%%%%
% To add notes in main text, please use \endnote{} and un-comment the codes below.
%\begin{adjustwidth}{-5.0cm}{0cm}
%\printendnotes[custom]
%\end{adjustwidth}
%%%%%%%%%%%%%%%%%%%%%%%%%%%%%%%%%%%%%%%%%%
\reftitle{References}


%\externalbibliography{yes}
%\bibliographystyle{JHEP}
%\bibliography{refs}
%=====================================
% References, variant B: internal bibliography
%=====================================

\providecommand{\href}[2]{#2}\begingroup\raggedright\endgroup

%\bibliographystyle{JHEP} %²Î¿¼ÎÄÏ×µÄ·ç¸ñ(.bst)
%\bibliography{refs} %²Î¿¼ÎÄÏ×ÎÄ¼þ(.bib)
\end{paracol}
% If authors have biography, please use the format below
%\section*{Short Biography of Authors}
%\bio
%{\raisebox{-0.35cm}{\includegraphics[width=3.5cm,height=5.3cm,clip,keepaspectratio]{Definitions/author1.pdf}}}
%{\textbf{Firstname Lastname} Biography of first author}
%
%\bio
%{\raisebox{-0.35cm}{\includegraphics[width=3.5cm,height=5.3cm,clip,keepaspectratio]{Definitions/author2.jpg}}}
%{\textbf{Firstname Lastname} Biography of second author}

% The following MDPI journals use author-date citation: Admsci,  Arts, Econometrics, Economies, Genealogy, Humanities, IJFS, Jintelligence, JRFM, Languages, Laws, Literature, Religions, Risks, Social Sciences. For those journals, please follow the formatting guidelines on http://www.mdpi.com/authors/references
% To cite two works by the same author: \citeauthor{ref-journal-1a} (\citeyear{ref-journal-1a}, \citeyear{ref-journal-1b}). This produces: Whittaker (1967, 1975)
% To cite two works by the same author with specific pages: \citeauthor{ref-journal-3a} (\citeyear{ref-journal-3a}, p. 328; \citeyear{ref-journal-3b}, p.475). This produces: Wong (1999, p. 328; 2000, p. 475)

%%%%%%%%%%%%%%%%%%%%%%%%%%%%%%%%%%%%%%%%%%
%% for journal Sci
%\reviewreports{\\
%Reviewer 1 comments and authors’ response\\
%Reviewer 2 comments and authors’ response\\
%Reviewer 3 comments and authors’ response
%}
%%%%%%%%%%%%%%%%%%%%%%%%%%%%%%%%%%%%%%%%%%
\end{document}